\documentclass[prl,aps,amsfonts,amssymb,draft,floats,twocolumn]{revtex4}

\newcommand{\bfn}{{\bf \vec n}}

\newcommand{\beg}{\begin{eqnarray}}
\newcommand{\eee}{\end{eqnarray}}
\newcommand{\f}{\frac}
\newcommand{\ccc}{{\vec{\sf C}}}
\newcommand{\be}{\begin{equation}}
\newcommand{\ee}{\end{equation}}

\begin{document}
\def\comment#1{}

\title{Reply to the comment by D. Gorokhov cond-mat/0502083}

\author{E. Babaev}
\affiliation{Laboratory for Atomic and Solid State Physics, Cornell University, Ithaca, NY 14853-2501, USA \\
Department of Physics, Norwegian University of
Science and Technology, N-7491 Trondheim, Norway}

\begin{abstract}We show that the recent comment by D. Gorokhov,
is based on physically obvious errors and misunderstandings of the content of the criticized papers
and is readily refuted.
We show that 1) In our paper [1] there were considered situations
of both strong and weak interband coupling regimes. 
2) In the ref [1] it was given a wide range
of physical systems with Josephson coupling strength ranging from very strong to
being exactly zero on symmetry grounds.
3)  While the ref [1]
is not a phenomenological study of $MgB_2$, the moderately strong Josephson coupling
results of [1] apply to $MgB_2$: the described in [1]
double-core vortices have been recently observed in $MgB_2$.

\end{abstract}
\maketitle

The first remark we would like to make is
that the Comment is unfortunately based on a misunderstanding of
a point which was  stressed in all our papers: that 
in two-band superconductors $U(1)\times U(1)$
symmetry is {\it strictly} forbidden because condensates
are not independently conserved and are 
coupled by Josephson term. The
BKT {\it transition} of the type discussed
in \cite{Babaev,mybkt} could not exist
in principle in two-band superconductors.
Specific  systems with a true $U(1) \times U(1)$ symmetry 
(i.e. without Josephson coupling) were
proposed in \cite{Babaev,mybkt}. 
A different story is indeed a 
possibility 
of occurrence in
in Josephson coupled systems of finite size BKT-like {\it crossovers} (a study
announced in \cite{Babaev} as the second paper in ref. [18]), we shall remark
on it below.

Another misunderstanding
on which the comment is
based is an assumption
that even
parts of the paper dealing with $U(1)\times U(1)$
symmetry and {\it zero} Josephson term were devoted to $MgB_2$.
In \cite{Babaev} we considered all general
situations ranging from condensates
with strong interband coupling to multicomponent
condensates where interband Josephson coupling
is forbidden. The paper \cite{Babaev} was not in any respect
a phenomenological study of $MgB_2$, 
($MgB_2$ was listed among many other examples
of weakly and strongly coupled two-gap
superconductors)
however
the  moderate Josephson-coupling results in \cite{Babaev}
are indeed relevant for $MgB_2$.
In particular 
double-core integer flux vortices (a linearly
bound state of two co-centered fractional vortices, i.e.
type-``({\it ii})" described on the page 3 of \cite{Babaev}) 
were indeed observed in $MgB_2$ \cite{Eskildsen}.
Therefore the potential applicability of the results of \cite{Babaev}
to $MgB_2$ mentioned in the abstract turned out to be correct.
When we considered zero or weak Josephson coupling
limits in \cite{Babaev,mybkt} we listed the systems where it is the case
like projected superconducting states of light atoms under extreme
pressure, certain states of spin-triplet superconductors
as well as Josephson-suppressed bilayer systems.

The technical side of the Comment is a
substitution of the well  known numbers characterizing
interband Josephson coupling from ref. \cite{Gurevich}
to the equations in \cite{Babaev}, all the equations in the comment
can be found in \cite{Babaev} but simply in a different
notation. 
Therefore nothing new in this respect  is revealed.  
That is, in particular Gorokhov asserts: { \sl
{(\bf A)}. ` finite coupling $g\neq 0$ generates a new length scale
$\Lambda$; for $R \agt \Lambda$ vortex--anti-vortex pairs attract
with a potential} {\it linear} { \sl in $R$ and thus exhibit confinement,
i.e., the BKT-transition is quenched.'
{(\bf B)}. `However, if $\Lambda$ is
much larger than the vortex core size, a BKT-like crossover
smeared on the scale $\Lambda$ can still be
observed', } {\bf Response for A)}:
The ``{\it linear}"  interaction of Josephson
vortices  has indeed been discussed in \cite{Babaev}:
the length scale $\Lambda$ has been
also discussed but merely in different notations
being called  ``the inverse mass 
for $n_1$ component of the unit vector
 $\vec{\bf n}$". The fact that in the presence of the Josephson effect we have sine-Gordon 
vortices is discussed in detail in the paper  (e.g. second page, left column).
 In particular it was written on the 
large Josephson coupling limit:
{\it ... the
energy per unit length  of noncomposite vortices 
is divergent in an infinite sample both in cases of zero and nonzero Josephson
coupling (in  case of finite $\eta$ 
a vortex creates a domain wall
 which makes
its energy per unit length divergent in infinite sample...}
{\bf Response for B)}:
The effect that even in a presence
of finite Josephson coupling 
there is  a length  scale where
the BKT transition-like crossover can be observed
is  also mentioned briefly in conclusion though
 conditions
for disappearance of the BKT transition
were not discussed because the case for
 finite-$\eta$ was reserved for a separate paper  
 (second paper cited as Ref.[13] in \cite{Babaev}):
%citing a ``future" paper ``Babaev and Gorokhov to be published":
{\it Moreover the BKT transition in a system of these vortices should be observable even
 in a type-I system 
both in the limits $\eta=0$,
and when   $\eta$ is large, where one has 
sine-Gordon vortices interacting with a linear potential [13]
(in the later case we apparently speak about a
finite size crossover).
}
Here we stress that Josephson coupling is a singular
perturbation any amount of it eliminates a true BKT transition,
a question of the observation of finite size-crossovers
in an experiment
is  more complicated than what was assumed in \cite{Gorokhov}
and depends on type of experimental probe and requires 
stricter criteria. Because we do not consider finite-size
crossovers of much interest, this question will not be
detailed here.

A remark on the point {\it i} in the comment:
As mentioned above in the paper \cite{Babaev}
we considered different limits, in particular
solutions for vortex in a general case of zero
Josephson coupling, also there
was given a criterion $L<\Lambda$ when 
in a general system (whether it is a superconductor
or layered system) Josephson coupling can be neglected in the
whole sample (where $L$ is the sample dimension
and $\Lambda$ is the Josephson length).
This sort of criteria is indeed applicable to
a sample which is large compared to other length scales
in the problem.
Systems where it is the case
were listed before going to this limit in \cite{Babaev}
with no $MgB_2$ mentioned in the list.

Also indeed  \cite{Babaev} does not feature absurd 
  statements that coherence,
penetration and Josephson lengths
``can be chosen arbitrarily for every superconductor"
and that ``one can fit vortices in sample smaller
than coherence length"
which were attributed in the Comment to 
\cite{Babaev} for unclear reasons.

We note that indeed \cite{Babaev}
was not a phenomenological
study of $MgB_2$ in any respect,
rather oppositely: albeit
essential physics of the Josephson
coupled superconductivity
discussed in \cite{Babaev} applies
to $MgB_2$, it is in fact one
of the least interesting applications
of the questions discussed in \cite{Babaev}.
We also remark that in recent years there appeared more physical
systems which were proposed either to  be
two-gap superconductors or nonsuperconducting
systems  where $U(1)\times U(1)$
symmetry appears as an effective description \cite{more}.

%A general remark regarding the value of Josephson coupling in $MgB_2$
%and two-band superconductors: while at the momen
In the Comment it is 
 also  claimed that the BKT transition
physics in \cite{Babaev,mybkt}
is ``well established" and experimentally observed
in layered systems ~\cite{Clem,Artemenko}.
First the similarities
with layered system physics were
discussed in \cite{Babaev,mybkt}, second
the differences between
layered (spatially separated) condensates
and two-gap superconductors are apparent, for example
the former case is not described by the
extended Faddeev model in \cite{Babaev}
because of the spatial separation. Third,
the flux carried by a vortex in one layer
in a system of $N$ identical layers \cite{Clem}
is a function of 
layer thickness, penetration length and distance
to surface. It should not be confused
with the flux quantization 
in a general two-gap superconductors
with arbitrary ratio of spatially
nonseparated condensates  given by eq. (5) in \cite{Babaev}.
 A more  important
circumstance is 
that the phase transition and the experimental
probe in ~\cite{Artemenko}
have little to do with the transition
considered in \cite{Babaev,mybkt}.
That is, we did not consider a  {\it superconducting}
transition, our point was a separation
of variables in general case and identification
of a state with quasi-long-range order
in phase difference which was discussed
explicitly in \cite{mybkt}. In layered
systems (connection to which was
indeed made in \cite{Babaev,mybkt}) such a transition is related
to dissipationless oppositely directed supercurrents belonging
to two layers. For layered superconductors
a proposal for concrete counterflow
experiment  and corresponding
calculations were done only this year \cite{DBG}
and no experimental confirmation 
of this transition has been yet reported.

%\newpage
{\bf Regarding the criticism of the experimental
paper by Festin et. al.}
 The  comment \cite{Gorokhov}
 also features criticism of the experimental paper by Festin et.al.
This discussion is also based on  physically obvious errors and 
in part on attribution to \cite{Festin} claims
which were not made there. 
 Festin et. al. never claimed that for Abrikosov vortices
BKT-like crossover 
(not a transition indeed) cannot be observed but in fact
they are the authors of a PRL paper where such
a crossover was observed Phys. Rev. Lett. 83, 5567 (1999). 
In the cond-mat/0303337 
the observation by Festin at. al. is different:
they  observed that a relative sharpness of crossover
in very thick $MgB_2$ films was unexpectedly much
narrower than that in much  thinner $YBCO$ films which 
reasonably led
to a possibility of very weakly coupled bands interpretation
in \cite{Festin}. 
 Later other experiments
and calculations gave opposite picture
about which the authors of \cite{Festin}
 were well informed long time ago.
It should be noted that the measurements
\cite{Festin} were done much earlier
than the publication of the eprint \cite{Festin}
and back then there  was no consensus on
interband coupling strength in $MgB_2$
and in particular there were reasons to
expect it being very weak.
Interband coupling can vary in a wide range
and even can be either  positive or negative.
Microscopic origin of possibility 
of weak interband couplings in two-band
system can be found in a number of
publications including \cite{Gurevich,Zhitomirsky}.
Detailed duiscussion of vortex physics 
in $MgB_2$ can also be found in e.g. \cite{Zhitomirsky}

The claim in the Comment that if
Abrikosov vortices have logarithmic
interaction at some finite scale
it results in a sharp BKT transition
is also based on a physically obvious error:
in a charged system a vortex has a finite-range interaction and
finite energy. Therefore  single vortices can be excited
by thermal fluctuations and an existence of a certain 
scale of logarithmic interaction does not lead
to a BKT transition. 
It is a well known
exact result that that for one-component system
 with a gauged $
U(1)$ symmetry there is no true sharp
 BKT transitions and no superfluid density jump. An existence of ill-defined 
BKT crossovers is indeed possible but 
that was not  denied by Festin et. al. In fact, as mentioned above such
questions were studied in their previous publications.
Besides that Festin et. al. studied granular samples
(which were essentially Josephson junctions arrays)
and any serious discussion of the multiple peaks
experiment \cite{Festin} should take into
accout this circumstance first of all.
%so the 

%In conclusion: the criticism in the Comment
%is based on an erroneous 
%interpretation of \cite{Babaev,mybkt}
%as a phenomenological study of $MgB_2$,
%(while a number of physical applications
%including situations with $U(1) \times U(1)$
%is proposed in \cite{Babaev}),
%reiteration of 
% known facts about interband coupling
%derived in $MgB_2$ \cite{Gurevich} (vortex
%physics in $MgB_2$ including interband
%coupling was discussed in great details in e.g. \cite{Zhitomirsky}).
% and attribution of claims and is essentially missing academic content.

On a separate note we would like to remark on a question
of observability of fractional flux in a situation of
 nonzero Josephson coupling. A very large ratio
of the coherence length to the Josephson length
noticed in \cite{Gorokhov} after substituting
numbers from \cite{Gurevich} to corresponding
equations is an apparent consequence of 
being extremely close to $T_c$ taking into
account temperature dependence of $\Lambda$
and $\xi$. This particular point not only does not
adequately characterizes strength of
interband coupling of any material in full
range of temperatures but 
it also does not invalidate 
a possibility to observe split
fractional vortices
in principle. Besides BKT transitions there is
a number of other possibilities to 
induce vortices, one such a possibility
is to exploit duality to Faddeev-Skyrme model
which is robust against Josephson term perturbation
and there are situations when Faddeev-Skyrme
term can provide a repulsive force between
two vortices with phase windings in
only one order parameter (details can be found
in Appendix)

Summary of points:

1. Gorokhov asserts that \cite{Babaev} is a study of 
$MgB_2$ and
fractional vortices in the limit of zero or weak Josephson
coupling. {Answer:} There were considered both limits
of weak and strong and zero Josephson coupling
in a general two-gap Ginzburg-Landau functional, it was shown that 
in strong Josephson coupling regime vortices
are confined linearly. Examples of a range of systems
with weak or zero Josephson coupling were given.

2. Albeit in \cite{Babaev} 
a phenomenological discussion of $MgB_2$
was not even attempted
however moderate coupling
limit considered in the paper is applicable
to $MgB_2$, in particular double-core integer
flux vortices described in \cite{Babaev}
were later observed in \cite{Eskildsen}.
{\it Therefore potential applicability to $MgB_2$ mentioned
in the abstract of \cite{Babaev} turned out to be correct.}

\appendix{\bf Appendix}
So the question is: if a Ginzburg-Landau model exists  with a mderately strong Josephson term
(e.g. just strong enough to forbid the BKT mechanism
 for thermal creation of pairs of fractional vortices),
could it nonetheless possess fractional vortices as spatially separated topological excitations?
The answer is positive: 
%in my papers in contrast to any previous
in our papers the variables were separated
in general case and in we have shown in \cite{we} that 
if to go beyond the London limit, two-gap
superconductor has a self-induced Faddeev-Skyrme term,
which counter-balances Josephson term in the circumstances discussed below:
 If we go beyond London limit and consider the order parameter $\bfn$ 
\cite{we},
we observe that the model also
admits ``baby" Skyrmions \cite{skyr} which are topological defects of the ${\bf R}^2 \to S^2$
map characterized by topological charge ${\rm deg}[\bfn] = 1/4\pi \int d^2 x \bfn \cdot \partial_1 \bfn \times 
\partial_2 \bfn $. The addition of mass terms 
like the Josephson term  $\rho^2 K n_1$ \cite{Babaev} is a 
{ necessary} condition
for the existence of stable baby skyrmions, which in the  absence of 
mass terms for $\bfn$ diverges \cite{skyr}  (there is also a mass term for $n_3$ coming
from Ginzburg-Landau potential \cite{we}). 
Despite in terms of the variable $\bfn$, a baby Skyrmion is a coreless
object, however the situation is actually more complicated
because the order parameter $\bfn=(\sin\theta\cos(\phi_1-\phi_2),\sin\theta\sin(\phi_1-\phi_2),\cos\theta)$
is defined with the help of the angle $\theta$ given by:
 $ |\Psi_{1,2}| = [ \rho \sqrt{2m_1} \sin (\f{\theta}{2}), 
\rho \sqrt{2m_2}  \cos(\f{\theta}{2}) ] $. Thus north and south poles of the order parameter space $S^2$  
correspond to zero of  the condensates $|\Psi_1|$ and $|\Psi_2|$  in physical space.
Thus a baby skyrmion in two-gap superconductor makes physical space multiply connected and one 
must impose singlevaluedness condition:  around zeroes of $|\Psi_{1,2}| $
the phases $\phi_{1,2}$ should change $2\pi$ times integer. In \cite{sk}
we  show that, for a defect with a given Hopf invariant, 
the winding of $(\phi_1-\phi_2)$, specified by the Hopf invariant, is consistent
 with singlevaluedness conditions only when
one has the following  phase windings around
these lines of zeroes: 
 $(\Delta \phi_i= 2\pi,\Delta \phi_j=0)$. This condition leads
to a nontrivial configuration of the field $\ccc$ \cite{sk};
{\it thus in a baby Skyrmion of ${\bf R}^2 \to S^2$ map,
 preimages of north and south poles of  $S^2$ are the fractional vortices}.
So, a baby Skyrmion in a TGS, in a simplest case emits two fractional
vortices like that considered in \cite{Babaev}.
These fractional vortices   attract each other;
however the attraction is  counterbalanced by the Faddeev-Skyrme term
which provides a repulsive force \cite{we}.

\end{document}